# Enhanced spin-to-charge conversion in La$_{0.67}$Sr$_{0.33}$MnO$_3$/NdNiO$_3$ bilayers at the nickelate metal-insulator phase transition

Biswajit Sahoo*, Sarmistha Das*, Akilan K, Alexandre Pofelski, Sebastien Petit-Watelot, Juan-Carlos Rojas-Sánchez, Yimei Zhu, Alex Frano[m], and Eric E Fullerton[m].

*These authors contributed equally to the work

B. Sahoo, E. E. Fullerton

Email: efullerton@ucsd.edu;

Center for Memory and Recording Research, University of California, San Diego, CA 92093, USA

B. Sahoo, A. Frano, S. Das

Department of Physics, University of California, San Diego, CA 92093, USA

Email: afrano@physics.ucsd.edu;

A. Pofelski, Y. Zhu

Brookhaven National Laboratory, 98 Rochester St, Upton, NY 11973, USA

Akilan K., S. Petit-Watelot, J-C. R. Sanchez

Université de Lorraine - CNRS (UMR 7198), Institut Jean Lamour, F-54011 Nancy Cedex, France

***Abstract:***

***Phase transition materials such as NdNiO$_3$ (NNO) when coupled with low damping ferromagnets such as La$_{0.67}$Sr$_{0.33}$MnO$_3$ (LSMO) can lead to new multi-functional material systems harnessing the interplay of charge, spin and orbital degrees of freedom. In this study, we probe the evolution of the spin-to-charge conversion in epitaxial all-oxide LSMO (12 nm)/NNO (4, 8, and 16 nm) bilayers. Using spin pumping ferromagnetic resonance we track the spin-charge conversion in the NNO layer through the paramagnetic metal to antiferromagnetic insulator transition and observe a pronounced enhancement of the inverse spin Hall effect signal at the onset of this transition. We attribute this enhancement to the electronic and magnetic disorder in NNO at the first-order phase transition, thereby providing insights into the mechanism of spin transport through the phase transition. The tunability of spin charge conversion in this low damping bilayer system offers a pathway for developing multifunctional, energy-efficient spintronic devices.***

Transition metal complex oxides offer a fertile playground to explore new physics and realize novel device functionalities owing to the correlated interaction of spin, charge, orbital and lattice degrees of freedom.[1–3] Rare earth nickelates, especially NdNiO$_3$ (NNO) have garnered special attention due to their first-order metal- insulator transition (MIT) (transition temperature T$_{MIT}$= 200 K in bulk), which is accompanied by interesting phenomena such as simultaneous change from paramagnetic to anti-ferromagnetic (AFM) ordering,[4–6] mixed phase coexistence with domain pattern formation, [7–11] and a structural phase transition.[5,12,13] Additionally, the MIT can be tuned by electron doping,[14–16] strain engineering,[17–26] thickness variation[27–29] and application of gate voltage.[30–32] While the control of MIT, by itself has potential application in development phase-transition based devices,[31,33] the multifunctionality of NNO can be further augmented by integrating with ferrimagnetic or ferromagnet layers. Interfacing NNO with a manganite ferromagnet layer has been shown to induce charge transfer effects [34] and proximity-induced novel ferromagnetism in NNO.[35–38] These interactions further enrich the phase diagram and provide a range of tuning mechanisms to control multiple degrees of freedom.

Here we report on the spin dynamics in epitaxial $La_{0.67}Sr_{0.33}MnO_3$ (LSMO) -NNO bilayers where we track the rather unexplored spin-to-charge conversion in NNO as it undergoes its MIT phase transition.[39] We observe thickness and temperature dependent modulation of spin-charge conversion within the NNO layer and observe an enhanced inverse spin Hall effect (ISHE) signal at the onset of the MIT. Our findings provide new insights into the interplay of the NNO phase transition with the LSMO magnetization dynamics, and its role in modulating spin currents. The tunable interactions can drive advancements in magnetism-based devices, such as magnetic tunnel junctions and spin-based nano-oscillators. The tunability of NNO, can modulate the spin dynamics in the ferromagnet, allowing for energy-efficient on-chip learning in neuromorphic hardware made from such interconnected devices.[40–45]

We studied LSMO/NNO bilayers of LSMO(12 nm)/NNO(0, 4, 8, and 16 nm) that were grown onto single-crystalline $NdGaO_3$(*110*) (NGO) substrates by pulsed laser deposition. The epitaxial orientation and film quality were evaluated by X-ray diffraction and transmission electron microscopy (TEM). The X-ray diffraction shows clear peaks of LSMO(*002*) and NNO(*220*), with Laue fringes indicating smooth films with excellent crystalline quality (Fig. S1(a)). The TEM imaging of a similarly grown LSMO(17 nm)/NNO(24 nm) bilayer (Fig.1) shows coherent growth of the LSMO/NNO bilayers, with a chemically sharp interface confirmed by electron energy loss spectroscopy (EELS) (Fig. S2). The electron diffraction pattern of NNO and LSMO films (Figs. 1(c) and (d)) further demonstrates the high crystalline quality of the films. NGO substrates were used as it has a low lattice mismatch with both LSMO (compressive) and NNO (tensile)and allows for growth of low damping LSMO layers [46,47] with a thickness dependent tunability of $T_{MIT}$ in NNO.[25]

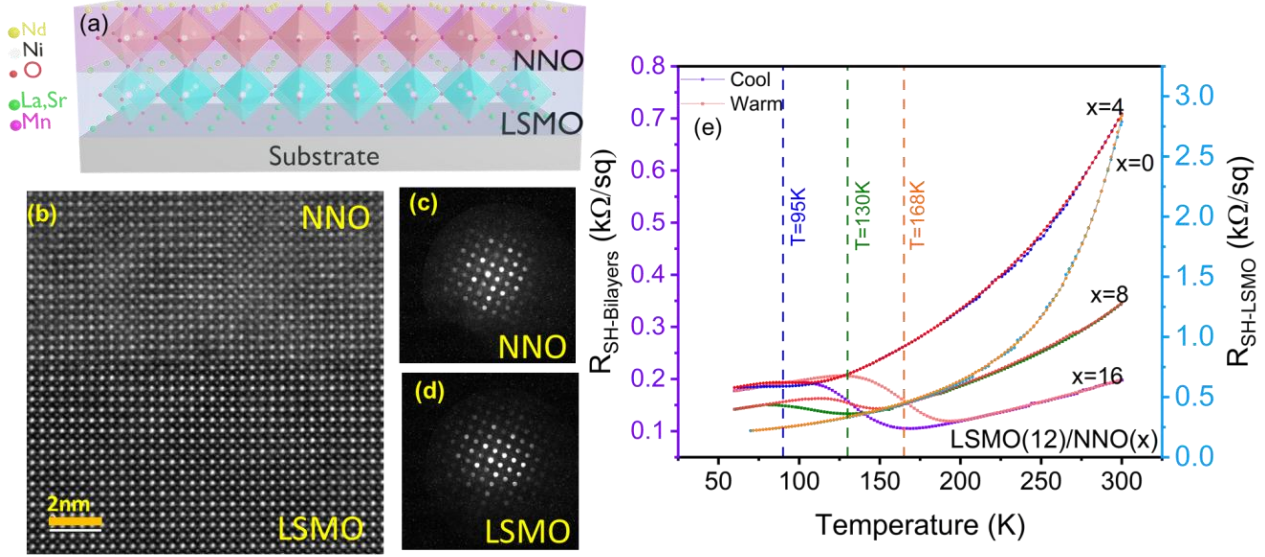

Figure 1 **Structural and electronic characterization:** (a) Shows a schematic of the epitaxial LSMO/NNO bilayer. (b) High-resolution cross-sectional TEM image of LSMO/NNO bilayer along the NGO<001> direction showing an atomically sharp interface with coherent growth of NNO on LSMO. (c) and (d) are the respective electron diffraction patterns. (e) Resistance vs temperature variation of samples LSMO(12)/NNO(0-16), with the $T_{MIT}$ in the cooling phase shown by dashed reference lines. The lower curves in each loop are from the cooling cycle.

Figure 1(e) shows the resistance ($R$) vs temperature ($T$) variation of the single-layer LSMO layer and bilayer films. The single LSMO layer shows the expected semi-metallic behavior for the FM phase of LSMO (Fig S3(f)), whereas the bilayers show the hysteretic behavior revealing the contribution from first-order MIT of NNO. The bilayer resistance including the hysteretic behavior closely follows a simple parallel resistor model of $R_{LSMO}$ and $R_{NNO}$ as shown in Fig S3(b). $T_{MIT}$ of NNO is 168K, 130K and 95K in the cooling cycle for the LSMO(12)/NNO(4, 8, 16) samples, respectively. Here $T_{MIT}$ is defined as the temperature marking the onset of the insulating behavior; *i.e.* where $dR/dT=0$ as outlined in Ref. 48 (Fig S3 (c)-(e)). We find that $T_{MIT}$ decreases with decreasing thickness of NNO.[18,25,26] This variation of $T_{MIT}$, involving only a few nanometers change in the thickness of the NNO layer while keeping the LSMO layer thickness constant, strongly supports the idea that the interfacial coupling between these two layers is inherently robust. Furthermore, the complex interfacial structural and charge interactions offer enhanced tunability, making this specific material combination highly suitable for our intended study. In bulk NNO, the MIT is concurrent with a magnetic transition from paramagnetic metal to an AFM insulating phase. However, in thin films the magnetic phase transition is observed to occur at a Néel temperature ($T_N$) somewhat lower than $T_{MIT}$[48] where there is correlated AFM order in the film. To estimate $T_N$ we measured the coercivity ($H_C$) of LSMO(12)/NNO(16) bilayer vs temperature and the $T_N$ was determined where $dH_C/dT$ changed with respect to temperature. For LSMO(12)/NNO(16) we estimate the $T_N$=130K in the cooling cycle (Fig S6(d)), which closely follows the $T_N$ determined from the peak of $d(\log(R))/d(1/T)$ vs T plots as outlined in Ref. 48 (Fig S3(c)-(e)). As mentioned before, we find $T_N < T_{MIT}$. All the above-mentioned observations for the bilayers derived from the electronic and magnetic properties are distinct from the intrinsic behaviors of the individual constituents of pristine LSMO and NNO. This strongly suggests the formation of a high-quality interface between the two materials as evidenced by TEM and XRD analyses and defined by $MnO_6$ and $NiO_6$ octahedral bonding. Moreover, these interfacial phenomena are sensitively modulated by the thickness of the NNO layer, further reinforcing the significance of this bilayer system for our study.[35–38] Thus, altogether having confirmed the interfacial magnetic coupling, the electronic response and the para-AFM transition + MIT in the bilayer films, we investigate the influence of the phase transition on the spin transport in NNO. In particular, we are interested in the spin-charge current conversion and its modulation thereof as NNO goes through its MIT.

We probe the spin-to-charge conversion in NNO via spin-pumping ferromagnetic resonance (SP-FMR) and fabricate micro-devices of 500 μm x 10 μm via a three step UV lithography (further details available in the **Methods** section). The device and measurement schematic are shown in Fig. 2(a-c). In SP-FMR, an external field ($H$) is simultaneously swept perpendicular to the radio frequency (RF) currents sent into the coplanar waveguides. The RF fields generated by the waveguide excite spin precession in the FM moments, which then pumps a pure spin current into the adjacent layer. This spin current is then converted into charge current by the inverse spin Hall effect (ISHE) and is probed by the voltage leads. The resulting lineshape is composed of a symmetric and anti-symmetric Lorentzian, with a voltage minimum or maximum at the resonant field ($H_{Res}$) of the FM, and is fit to the following equation:

$$V_{total} = V_S \frac{\left(\frac{\Delta H}{2}\right)^2}{\left(\frac{\Delta H}{2}\right)^2 + (H - H_{res})^2} + V_A \frac{\left(\frac{\Delta H}{2}\right)(H - H_{res})}{\left(\frac{\Delta H}{2}\right)^2 + (H - H_{res})^2} \tag{1}$$

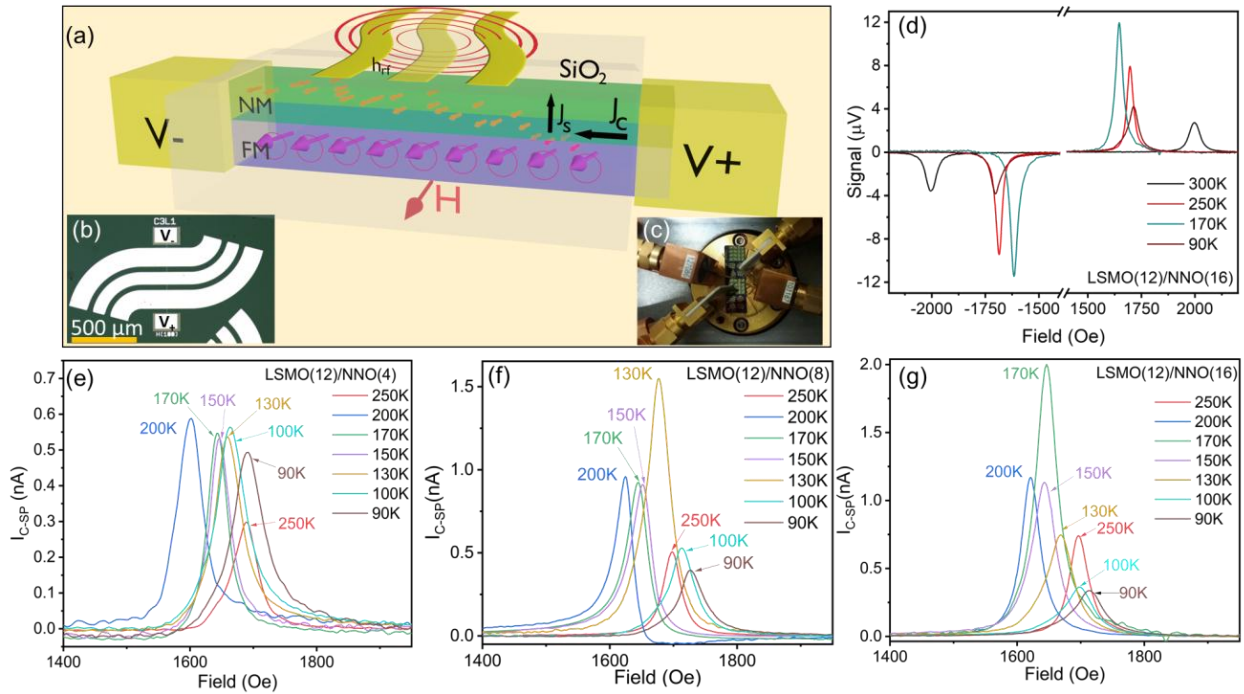

*Figure 2. **Transport and spin-pumping on LSMO/NNO**. (a)-(c) show the SP-FMR schematic with an optical micrograph of the device (b) and the probe placement (c). The top left and bottom right are the RF probes, the other two being the voltage probes. (d) shows the full-field sweep SP-FMR signal for LSMO(12)/NNO(16) at 8GHz and 15dBm power, demonstrating a dominant spin-pumping contribution. (e)-(g) show the spin pumped charge current ($I_{C-SP}$) for*

*LSMO(12)/NNO(4-16) at 8GHz at different temperatures in the cooling cycle demonstrating a distinct increase in the current at the $T_{MIT}$ for LSMO(12)/NNO(8-16). Measurements are performed with H//<1-10> direction of NGO(110) substrate. The $H_{res}$ reduces with temperature, increasing again for T< 200K, is due to magneto crystalline anisotropy of LSMO (see Supp. Info. )*

Here $\Delta H$ is the linewidth, $V_S$ is the symmetric contribution of the voltage which has major contributions from spin-pumping and minor contributions, if any, from planar Hall effect (PHE) and anisotropic magnetoresistance (AMR). $V_A$ is the anti-symmetric part, which consists of rectification effect dominated by anomalous Hall effect (AHE) . When spin-pumping is the dominant contribution then $V_S$ flips sign for positive and negative field sweeps, whereas the sign does not change for the AMR dominated contribution.[49] Single layer NGO(*110*)/LSMO samples do not show spin-pumping dominated behavior.[47]

Figure 2(d) shows the temperature dependent SP-FMR voltage signal of LSMO(12)/NNO(16) at 8GHz in the cooling cycle for some temperatures (90K<T<300K), demonstrating a strong spin-pumping dominated signal exemplified by the symmetric lineshape which flips sign upon reversing the magnetic field (full scans for all temperatures are plotted in the supplementary section). This is the case for LSMO(12)/NNO(4) and LSMO(12)/NNO(8) as well (full field sweeps are not shown for brevity).  In conclusion, the observation of this effect in all bilayers, including the one with the lowest NNO thickness of 4 nm, and its absence in single-layer LSMO grown on NGO(*110*) [47] clearly indicates that the NNO layer plays a crucial role for the spin-to-charge conversion. The sign of spin-charge conversion for NNO is the same as Pt[47] which is corroborated by the SP-FMR of a Py/NNO bilayer at RT as shown in Fig. S10 (a). Unlike metallic systems, the resistance of LSMO/NNO bilayers has a significant temperature variation (Fig. 1(e)) and thus, we evaluate the spin-pumped charge current ($I_{C-SP}=V_{SP}/R_{dev}$) for the cooling cycle. Figures 2(d)-(f) show the positive field sweep of the $I_{C-SP}$ for LSMO(12)/NNO(4-16). LSMO(12)/NNO(4) with a $T_{MIT}$=95K is expected to be metallic for 95K<T<300K. In this bilayer the $I_{C-SP}$ increases from 300K down to 200K, which can be attributed to the increased spin-current pumped by LSMO due to increased $M_s$ and exchange interaction strength of LSMO with decreasing temperature. The $I_{C-SP}$ shows a small reduction for T<200K, remains mostly constant until 130K and then decreases slightly again for T<130K. Such behavior has been observed in LSMO and LSMO/Pt system, which has been attributed to dissipation of spin currents in the bulk and interface of LSMO for *T*<130K.[47]

The more interesting variation happens for LSMO(12)/NNO(16) and LSMO(12)/NNO(8) where $T_{MIT}$≥ 130 K. For LSMO(12)/NNO(16), $I_{C-SP}$ increases from 300K to 200K with a significant (2-fold) enhancement at 170K, which is close to $T_{MIT}$ of this bilayer, before dropping to much smaller values for lower temperatures. We observe a similar trend in the $I_{C-SP}$ for LSMO(12)/NNO(8): $I_{C-SP}$ increases until 200K and reduces slightly and remains mostly constant for 170K<*T*<150K, (in accordance with what was observed for LSMO(12)/NNO(4)), and then shows significant increase at 130K ($T_{MIT}$ for LSMO(12)/NNO(8)), before reducing drastically for 90K<T<130K. The variation of $I_{C-SP}$ for LSMO(12)/NNO(4-16) with respect to the reduced temperature $T/T_{MIT}$ are plotted in Fig.3(a) ( with respect to *T* in supplementary Fig. S11(a)). As can be seen, there is a sharp peak in $I_{C-SP}$ at $T_{MIT}$ for the 8 and 16 nm NNO layers followed by a steep fall off for $T<T_{MIT}$. This is our key experimental result, a direct linking of the spin pumping characteristics to the MIT phase transition in NNO.

We can further estimate the efficiency of the conversion of spin current to charge current, the spin Hall angle ($\theta_{SHA}$), via the following relation[50–52]:

$$I_{C-SP} = \theta_{SHA} J_s w \lambda_{sf} \tanh\left(\frac{t_{NNO}}{2\lambda_{sf}}\right), \tag{2}$$

where $t_{NNO}$ is the thickness of NNO layer, $w$ is the width of the device (10 μm) and $\lambda_{SF}$ is the spin diffusion length of NNO. $\lambda_{SF}$ was determined by the exponential fitting of the Gilbert damping parameter with NNO thickness[53]. We evaluated the spin diffusion length in the metallic regime ($\lambda_{SF-M}$)170K<$T$≤300K to be ≈ 4.1 nm and in the insulating regime 90<=T<=100 to be $\lambda_{SF-I}$≈ 2.2 nm (See supplementary for details). (Fig. S9(a)-(b)). $J_s$ is the spin current density injected at the LSMO/NNO interface and it is determined by the following relation[50–52,54]:

$$J_s \approx \left(\frac{g_{eff}^{\uparrow\downarrow}\hbar}{8\pi}\right)\left(\frac{h_{RF}\gamma}{\alpha}\right)^2 \left[\frac{4\pi M_{eff}\gamma + \sqrt{(4\pi M_{eff}\gamma)^2 + (4\pi f)^2}}{(4\pi M_{eff}\gamma)^2 + (4\pi f)^2}\right]\left(\frac{2e}{\hbar}\right). \tag{3}$$

Here $e$ is the electronic charge and $h_{RF}$ is the RF field experienced by the sample which was calibrated beforehand for all RF frequencies[55], $\alpha$ is the Gilbert damping (obtained via FMR) and $g_{eff}$ is the effective spin mixing conductance given as $g_{eff}^{\uparrow\downarrow} = \frac{4\pi M_s d_{FM} \Delta\alpha}{g\mu_B}$,[56] where $g$ is the electronic g-factor, $\mu_B$ is the Bohr magneton. $\Delta\alpha$ is the difference between the damping of bilayer samples and the LSMO film given by $\alpha_{bilayer} - \alpha_{LSMO}$ obtained by FMR, $\alpha_{LSMO} = 0.003$, and $d_{FM}$ is the LSMO thickness We observe that above MIT, the overall efficiency increases with the NNO thickness following the reduction of its sheet resistance (Fig. 1e). This happens in systems when Dyakonov-Perels scattering is the dominant mechanism for spin relaxation.[57,58] The values of SHA range from 1-3% for T>200K, reaches a maximum close to MIT temperature and decreases drastically below this characteristic temperature due to insulator behavior of NNO; but SP-FMR signal still it might be observed due to some metallic-like system at LSMO/NNO interface (note: it has been shown for YIG/Pt[59] and Bi:YIG/Pt for instance, here we have the opposite, *i.e.* FM metallic and NM "insulator").[35,60] Another possible contributing factor may be the coherent interfacial bonding facilitated by oxygen octahedra sharing within the perovskite structure at the interface between NNO and LSMO. This interface enables direct interaction between two distinct magnetic layers, namely the ferromagnetic LSMO and the antiferromagnetic NNO, which gives rise to a random magnetic field due to interfacial coupling. This field is further influenced by the application of an external magnetic field and competes with other intrinsic energy contributions in the system, such as magnetic exchange interactions, magnetocrystalline anisotropy, and magnetostatic energy etc. As a result, these bilayers exhibit behavior that is not only distinct from their individual components: the pristine NNO and LSMO, but also from other conventional other metallic heterostructures. This distinct behavior extends beyond the metal-insulator transition observed in pristine NNO. The perovskite structure, with its corner-sharing octahedral framework, facilitates a strain-mediated structural continuity across the interface, allowing for dynamic octahedral rotations and distortions that further enhance interfacial coupling and such emergent phenomena.

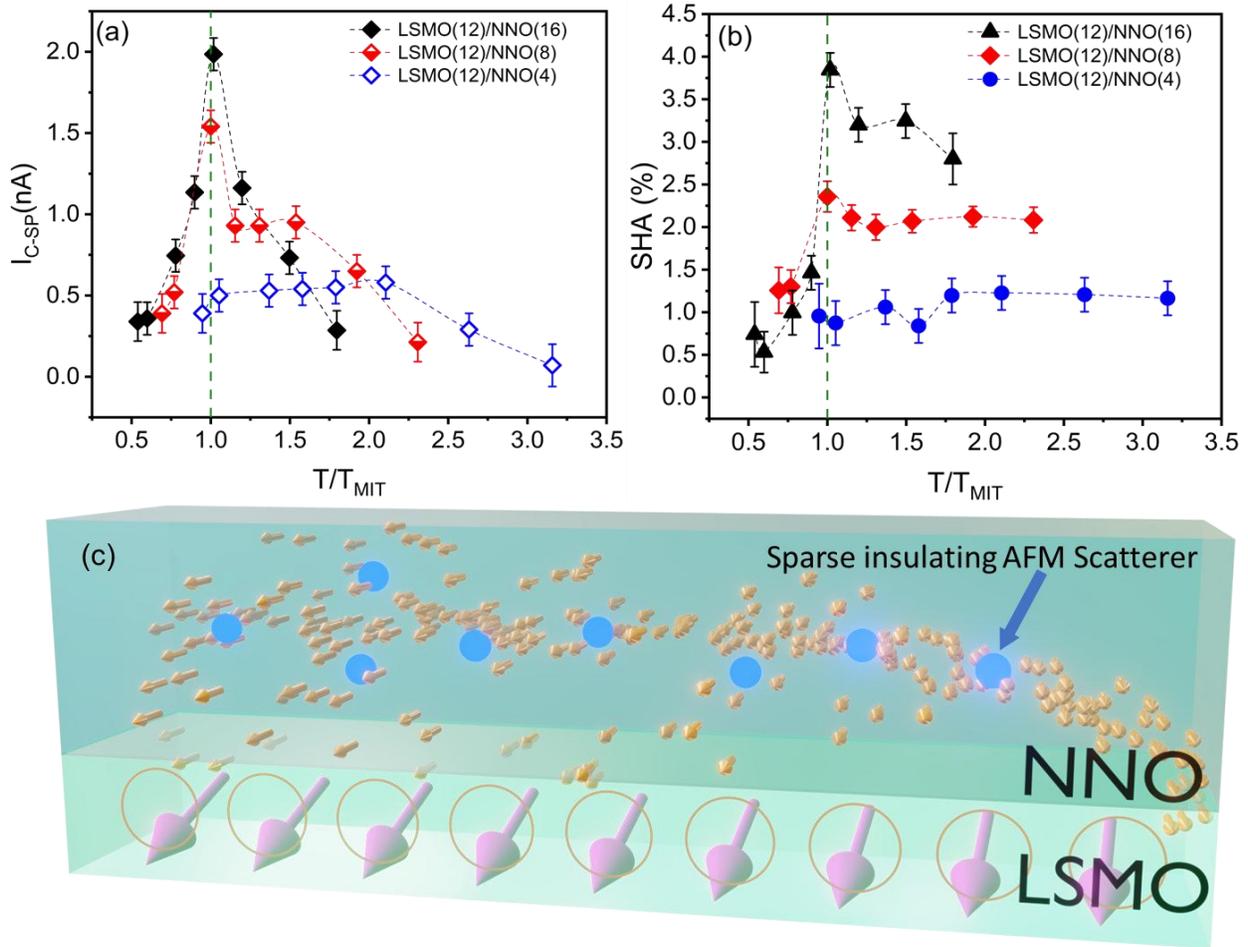

Figure 3. **Signatures of enhanced spin pumping MIT transition**. *(a) charge current produced by NNO layer pumped from LSMO at 8 GHz as function of normalized temperature $T/T_{MIT}$ for different NNO thickness. (b) Spin Hall angle variation with respect to $T/T_{MIT}$. We see a distinct peak when the temperature reaches $T_{MIT}$ (vertical green dotted reference line).(c) shows a schematic of the observed behavior, where the sparse insulating domains enhance the spin-charge conversion due to additional scattering. The orange arrows represent the pumped spins into NNO.*

The drastic reduction of current (188% and 76%) and the $\theta_{SHA}$ (81% and 162%) for LSMO(12)/NNO(8) and LSMO(12)/NNO(16), respectively, from the metallic to the insulating phase within 20K below $T_{MIT}$ can be attributed to the reduction of itinerant electrons in the insulating phase of NNO, which no longer efficiently convert the spin current into charge current. This is a well-known phenomenon to occur in the electron hopping/ dirty metal regime[59,61] and is further exemplified by the inverse variation of the spin Hall conductivity with the longitudinal resistivity (Fig. S11(c)). Additionally, the metallic NNO exhibits a larger $\theta_{SHA}$ for increased layer thicknesses, indicating a significant bulk contribution. Based on these observations, we conclude that the spin scattering mechanism in NNO operates in the dirty metal regime and is dominated by extrinsic contributions.

The drastic enhancement of $I_{C-SP}$ (and the $\theta_{SHA}$) at the onset of the MIT may be explained by the presence of additional extrinsic spin scatterers. The MIT of NNO occurs via nucleation of insulating domains, which start forming at the onset of MIT.[7,8] These abruptly formed sparse AFM insulating domains within a much larger metallic matrix can act as extrinsic spin scatterers leading to enhanced spin scattering with a consequent reduction in the scattering lifetime $\tau$. Availability of enough itinerant electrons can convert it into a large charge current, resulting in a much larger spin-charge conversion. It is known that the spin-charge conversion efficiency can be enhanced solely by reducing $\tau$ up to a certain limit.[62] This has been demonstrated in the case of Pt doped with insulating impurities such as MgO or TiO$_2$ resulting in increased spin-charge conversion efficiency with doping concentration as less as ≤ 3%.[62–64] In the case of NNO, the phase transition occurs by forming sparse domains that are both insulating and AFM[11], and here these insulating domains act as "doped" sparse impurities, similar to doped Pt systems.[62–64] These domains nucleate at $T_{MIT}$ and keep growing until they completely occupy the NNO film. For T< $T_{MIT}$ the insulating regions quickly take over[7], significantly decreasing $\tau$ and reducing the available itinerant electrons, subsequently causing a large reduction in the $I_{C-SP}$ and $\theta_{SHA}$. Such change in the charge to spin conversion is fundamentally different from that in other phase transition materials such as VO$_2$ in the sense that the domain formation in the latter is accompanied by a sharp structural phase change via filament formation[65,66] with no magnetic ordering, and does not seem to enhance the $I_{C-SP}$ (or $\theta_{SHA}$) at $T_{MIT}$.[59] We thus posit that the AFM nature of the domains may be further aiding in the spin dependent scattering of the spin current in NNO. In such systems like NNO/LSMO, the interplay of correlated electron behaviour, magnetic exchange across atomically sharp and structurally coherent interfaces, and strain-tunable octahedral connectivity introduces complex interfacial physics not accessible in metallic systems. Here, the spin-to-charge conversion is strongly influenced by intrinsic properties of the transition metal oxides, including their strongly coupled spin, orbital, and lattice degrees of freedom enabled by the continuity of the perovskite lattice and modulated by epitaxial strain. These emergent behaviours, driven by structural and magnetic reconstructions at the interface, go beyond conventional spin Hall or proximity-induced effects and point to an entirely different route for realizing tunable spin conversion phenomena in complex oxide systems.

In conclusion, we describe the first experimental results of the spin-to-charge conversion of NNO in epitaxial all-oxide LSMO/NNO bilayers. The spin current (and spin-charge conversion efficiency) modulation is primarily controlled by the MIT and thus can be furthered altered by controlling the phase transition of NNO via voltage induced strain(e.g. using PMN-Pt as substrate[23,67,68]), and resistance manipulation via gate voltage.[31] The tunable spin conduction of NNO modulated primarily by the MIT can allow it to act as a weighted spin valve in tri-layer systems e.g. NNO/FM/high-SOC-NM allowing controlled phase dependent manipulation of spin currents into the FM. NNO may also allow magnon based spin-transport in tri-layer systems[69] of high-SOC-NM /NNO/LSMO owing to the low damping and robust AFM coupling. This can bring forth multi-functional spintronic devices that can enable realization of devices such as spintronic hardware with on-chip learning for neuromorphic computing.

**Methods:**

**Sample Growth:** 12-nm-thick LSMO films were grown on NdGaO$_3$(*110*) (NGO) substrates by pulsed laser deposition. The substrate temperature during deposition was kept constant at 950$^0$C. The films were grown at an O$_2$ pressure of 200 mTorr. The films were then cooled down at the same pressure till 710$^0$C after which NNO was deposited at the same deposition pressure. The films were then annealed at the same temperature for 15 minutes at 7.6 Torr $O_2$ pressure and then cooled down to room temperature at the same pressure.

**Electrical, Magnetic and FMR characterization:** The RT measurement of the samples were performed via the 4 probe Van der Pauw method from 300K to 50K at 2K intervals in the warming and cooling cycles to obtain the sheet resistance. The resistance was measured with 5 averages after each temperature point was stabilized. Temperature dependent magnetometry was performed using a vibrating sample magnetometer (VSM) by Quantum Design to confirm the ferromagnetic phase of the samples. A field modulated flip chip coplanar waveguide FMR modified to adapt to the Quantum Design cryogenic chamber for temperature variation was used for FMR characterization.

**Device Fab:** For Fabricating SP-FMR devices, the film was patterned into devices of 500 x 10 µm wires via UV lithography and Ar ion etching. Next, 200 nm of $SiO_2$ was deposited on the entire substrate, leaving areas for gold contact deposition. Finally, 5-nm Ti and 150-nm Au layers were deposited by evaporation to fabricate ground-signal-ground waveguides and the contacts by lift-off. For measurements, the devices were placed in a cryo-chamber with 4 probes: two for sending RF currents and 50 Ohm termination; and two for measuring the resulting voltage. An external field perpendicular to the RF current was swept, and the resulting voltage (due to spin rectification effects) was recorded by a nano-voltmeter.

**STEM Characterization:** Samples were prepared using the in-situ lift-out method in the Helios G5 Dual Beam instrument operating at 30 keV. A final clean-up at 5 keV was performed to remove part of the amorphous material on the lamella. Samples were observed in a double aberration corrected JEOL ARM 200F cold FEG microscope operating at 200 keV. The STEM High Angle Annular Dark-Field (HAADF) imaging was done with a semi-convergence angle of 21 mrad and a current of 60 pA resulting in a probe size of approximately 80 pm. The Electron Energy Loss Spectroscopy (EELS) analysis was performed using the GIF Continuum spectrometer and the K3 direct detection camera in Dual EELS counting mode using a dwell time of 50 ms (5 ms and 45 ms for the low loss and core loss part, respectively). The energy alignment was performed using the zero-loss peak. The acquired core loss EELS spectra were denoised using the Principal Component Analysis method (PCA), keeping the first 20 components.

## Acknowledgements:

This work was primarily supported as part of Quantum Materials for Energy Efficient Neuromorphic Computing (Q-MEEN-C), an Energy Frontier Research Center funded by the U.S. Department of Energy (DOE), Office of Science, Basic Energy Sciences (BES), under Award # DE-SC0019273.

This work was performed in part at the San Diego Nanotechnology Infrastructure (SDNI) of UCSD, a member of the National Nanotechnology Coordinated Infrastructure, which is supported by the National Science Foundation (Grant ECCS-2025752).

The electron microscopy facility at the Brookhaven National Laboratory was supported by DOE-BES, Materials Science and Engineering Division, under Contract No. DE-SC0012704.

This material is based upon research partially supported by the Chateaubriand Fellowship of the Office for Science & Technology of the Embassy of France in the United States.

Devices were patterned at Jean Lamour Institute's cleanroom platform, MiNaLor, which is partially funded by the Grand Est region via the project RANGE. We also acknowledge the support from EU-H2020-RISE project Ultra-Thin Magneto Thermal Sensing ULTIMATE-I (Grant ID. 101007825). This result is also part of a project that has received funding from the European Research Council (ERC) under the European Union's Horizon 2020 research and innovation programme, Grant agreement No 101086807 (MAGNETALLIEN).

This work was partially supported by the project "Lorraine Université d'Excellence" reference ANR-15-IDEX-04-LUE, through the France 2030 government grants EMCOM (ANR-22-PEEL-0009), and PEPR SPIN – SPINMAT ANR-22-EXSP-0007. We thank Melissa Yactayo for complementary spin pumping measurements.

# References:

1. Imada, M., Fujimori, A. & Tokura, Y. Metal-insulator transitions. *Rev. Mod. Phys.* **70**, 1039–1263 (1998).

2. Tsukazaki, A. *et al.* Quantum Hall Effect in Polar Oxide Heterostructures. *Science* **315**, 1388–1391 (2007).

3. Salluzzo, M. *et al.* Orbital Reconstruction and the Two-Dimensional Electron Gas at the $LaAlO_3/SrTiO_3$ Interface. *Phys. Rev. Lett.* **102**, 166804 (2009).

4. Torrance, J. B., Lacorre, P., Nazzal, A. I., Ansaldo, E. J. & Niedermayer, Ch. Systematic study of insulator-metal transitions in perovskites $RNiO_3$ (R=Pr,Nd,Sm,Eu) due to closing of charge-transfer gap. *Phys. Rev. B* **45**, 8209–8212 (1992).

5. Medarde, M. L. Structural, magnetic and electronic properties of perovskites (R = rare earth). *J. Phys. Condens. Matter* **9**, 1679 (1997).

6. Bodenthin, Y. *et al.* Magnetic and electronic properties of RNiO3 (R = Pr, Nd, Eu, Ho and Y) perovskites studied by resonant soft x-ray magnetic powder diffraction. *J. Phys. Condens. Matter* **23**, 036002 (2011).

7. Mattoni, G. *et al.* Striped nanoscale phase separation at the metal–insulator transition of heteroepitaxial nickelates. *Nat. Commun.* **7**, 13141 (2016).

8. Preziosi, D. *et al.* Direct Mapping of Phase Separation across the Metal–Insulator Transition of NdNiO3. *Nano Lett.* **18**, 2226–2232 (2018).

9. Li, J. *et al.* Scale-invariant magnetic textures in the strongly correlated oxide $NdNiO_3$. *Nat. Commun.* **10**, 4568 (2019).

10. Bisht, R. S., Samanta, S. & Raychaudhuri, A. K. Phase coexistence near the metal-insulator transition in a compressively strained $NdNiO_3$ film grown on $LaAlO_3$: Scanning tunneling, noise, and impedance spectroscopy studies. *Phys. Rev. B* **95**, 115147 (2017).

11. Bluschke, M. *et al.* Imaging mesoscopic antiferromagnetic spin textures in the dilute limit from single-geometry resonant coherent x-ray diffraction. *Sci. Adv.* **8**, eabn6882 (2022).

12. Mazin, I. I. *et al.* Charge Ordering as Alternative to Jahn-Teller Distortion. *Phys. Rev. Lett.* **98**, 176406 (2007).

13. García-Muñoz, J. L., Aranda, M. A. G., Alonso, J. A. & Martínez-Lope, M. J. Structure and charge order in the antiferromagnetic band-insulating phase of NdNiO$_3$. *Phys. Rev. B* **79**, 134432 (2009).

14. Sidik, U., Hattori, A. N., Rakshit, R., Ramanathan, S. & Tanaka, H. Catalytic Hydrogen Doping of NdNiO$_3$ Thin Films under Electric Fields. *ACS Appl. Mater. Interfaces* **12**, 54955–54962 (2020).

15. Ren, H. *et al.* Controllable Strongly Electron-Correlated Properties of NdNiO$_3$ Induced by Large-Area Protonation with Metal–Acid Treatment. *ACS Appl. Electron. Mater.* **4**, 3495–3502 (2022).

16. Song, Q. *et al.* Antiferromagnetic metal phase in an electron-doped rare-earth nickelate. *Nat. Phys.* **19**, 522–528 (2023).

17. Catalan, G., Bowman, R. M. & Gregg, J. M. Metal-insulator transitions in NdNiO$_3$ thin films. *Phys. Rev. B* **62**, 7892–7900 (2000).

18. Tiwari, A., Jin, C. & Narayan, J. Strain-induced tuning of metal–insulator transition in NdNiO3. *Appl. Phys. Lett.* **80**, 4039–4041 (2002).

19. Kumar, Y., Choudhary, R. J. & Kumar, R. Strain controlled systematic variation of metal-insulator transition in epitaxial NdNiO$_3$ thin films. *J. Appl. Phys.* **112**, 073718 (2012).

20. Liu, J. *et al.* Heterointerface engineered electronic and magnetic phases of NdNiO$_3$ thin films. *Nat. Commun.* **4**, 2714 (2013).

21. Mikheev, E. *et al.* Tuning bad metal and non-Fermi liquid behavior in a Mott material: Rare-earth nickelate thin films. *Sci. Adv.* **1**, e1500797 (2015).

22. Hauser, A. J. *et al.* Correlation between stoichiometry, strain, and metal-insulator transitions of NdNiO3 films. *Appl. Phys. Lett.* **106**, 092104 (2015).

23. Heo, S. *et al.* Modulation of metal-insulator transitions by field-controlled strain in NdNiO3/SrTiO3/PMN-PT (001) heterostructures. *Sci. Rep.* **6**, 22228 (2016).

24. Yao, D. *et al.* Tuning the metal-insulator transition via epitaxial strain and Co doping in NdNiO3 thin films grown by polymer-assisted deposition. *J. Appl. Phys.* **119**, 035303 (2016).

25. Zhang, J. Y., Kim, H., Mikheev, E., Hauser, A. J. & Stemmer, S. Key role of lattice symmetry in the metal-insulator transition of NdNiO3 films. *Sci. Rep.* **6**, 23652 (2016).

26. Guo, Q., Farokhipoor, S., Magén, C., Rivadulla, F. & Noheda, B. Tunable resistivity exponents in the metallic phase of epitaxial nickelates. *Nat. Commun.* **11**, 2949 (2020).

27. Laffez, P. *et al.* Evidence of strain induced structural change in hetero-epitaxial NdNiO$_3$ thin films with metal-insulator transition. *Eur. Phys. J. Appl. Phys.* **25**, 25–31 (2004).

28. Palina, N. *et al.* Investigation of the metal–insulator transition in NdNiO3 films by site-selective X-ray absorption spectroscopy. *Nanoscale* **9**, 6094–6102 (2017).

29. Suyolcu, Y. E. *et al*. Control of the metal-insulator transition in NdNiO$_3$ thin films through the interplay between structural and electronic properties. *Phys. Rev. Mater.* **5**, 045001 (2021).

30. Dong, Y. *et al*. Effect of gate voltage polarity on the ionic liquid gating behavior of NdNiO$_3$/NdGaO$_3$ heterostructures. *APL Mater.* **5**, 051101 (2017).

31. Scherwitzl, R. *et al*. Electric-Field Control of the Metal-Insulator Transition in Ultrathin NdNiO$_3$ Films. *Adv. Mater.* **22**, 5517–5520 (2010).

32. Asanuma, S. *et al*. Tuning of the metal-insulator transition in electrolyte-gated NdNiO3 thin films. *Appl. Phys. Lett.* **97**, 142110 (2010).

33. Zhao, J. *et al*. Memristors based on NdNiO$_3$ nanocrystals film as sensory neurons for neuromorphic computing. *Mater. Horiz*. **10**, 4521–4531 (2023).

34. Xu, Z. *et al*. Strain-Enhanced Charge Transfer and Magnetism at a Manganite/Nickelate Interface. *ACS Appl. Mater. Interfaces* **10**, 30803–30810 (2018).

35. Das, S., Prajapati, G. L., Santhosh Kumar, K. & Rana, D. S. Interface exchange coupling controlled electronic transport in (La,Nd)NiO3/La0.7Sr0.3MnO3 multilayer thin films. *J. Magn. Magn. Mater.* **490**, 165487 (2019).

36. Pandey, P., Rana, R., Tripathi, S. & Rana, D. S. Effect of structural and magnetic exchange coupling on the electronic transport of NdNiO3 films intercalated with La0.7Sr0.3MnO3 thin layers. *Appl. Phys. Lett.* **103**, 032403 (2013).

37. Chen, K. *et al*. Charge-transfer-induced interfacial ferromagnetism in La$_{0.7}$Sr$_{0.3}$MnO$_3$/NdNiO$_3$. *Phys. Rev. Mater.* **4**, 054408 (2020).

38. Caputo, M. *et al*. Proximity-Induced Novel Ferromagnetism Accompanied with Resolute Metallicity in NdNiO3 Heterostructure. *Adv. Sci.* **8**, 2101516 (2021).

39. Jeong, S. *et al*. Enhanced spin–orbit torque in Ni$_{81}$Fe$_{19}$/Pt bilayer with NdNiO$_3$ contact. *Appl. Phys. Lett.* **119**, 212402 (2021).

40. Böhnert, T. *et al*. Weighted spin torque nano-oscillator system for neuromorphic computing. *Commun. Eng.* **2**, 1–8 (2023).

41. Romera, M. *et al*. Binding events through the mutual synchronization of spintronic nano-neurons. *Nat. Commun.* **13**, 883 (2022).

42. Vodenicarevic, D., Locatelli, N., Grollier, J. & Querlioz, D. Nano-oscillator-based classification with a machine learning-compatible architecture. *J. Appl. Phys.* **124**, 152117 (2018).

43. Zahedinejad, M. *et al*. Two-dimensional mutually synchronized spin Hall nano-oscillator arrays for neuromorphic computing. *Nat. Nanotechnol.* **15**, 47–52 (2020).

44. Zahedinejad, M. *et al*. Memristive control of mutual spin Hall nano-oscillator synchronization for neuromorphic computing. *Nat. Mater.* **21**, 81–87 (2022).

45. Zhang, H.-T. *et al*. Reconfigurable perovskite nickelate electronics for artificial intelligence. *Science* **375**, 533–539 (2022).

46. Qin, Q. *et al*. Ultra-low magnetic damping of perovskite La0.7Sr0.3MnO3 thin films. *Appl. Phys. Lett*. **110**, 112401 (2017).

47. Sahoo, B. *et al*. Temperature Dependent Spin Dynamics in $La_{0.67}Sr_{0.33}MnO_3$/Pt Bilayers. *Adv. Mater. Interfaces* **12**, 2401038 (2025).

48. Ojha, S. K. *et al*. Anomalous electron transport in epitaxial $NdNiO_3$ films. *Phys. Rev. B* **99**, 235153 (2019).

49. Harder, M., Gui, Y. & Hu, C.-M. Electrical detection of magnetization dynamics via spin rectification effects. *Phys. Rep*. **661**, 1–59 (2016).

50. Sahoo, B. *et al*. Spin Pumping and Inverse Spin Hall Effect in Iridium Oxide. *Adv. Quantum Technol*. **4**, 2000146 (2021).

51. Rogdakis, K. *et al*. Spin transport parameters of NbN thin films characterized by spin pumping experiments. *Phys. Rev. Mater*. **3**, 014406 (2019).

52. Rojas-Sánchez, J.-C. *et al*. Spin Pumping and Inverse Spin Hall Effect in Platinum: The Essential Role of Spin-Memory Loss at Metallic Interfaces. *Phys. Rev. Lett*. **112**, 106602 (2014).

53. Fache, T., Rojas-Sanchez, J. C., Badie, L., Mangin, S. & Petit-Watelot, S. Determination of spin Hall angle, spin mixing conductance, and spin diffusion length in CoFeB/Ir for spin-orbitronic devices. *Phys. Rev. B* **102**, 064425 (2020).

54. Ando, K. & Saitoh, E. Inverse spin-Hall effect in palladium at room temperature. *J. Appl. Phys*. **108**, 113925 (2010).

55. Anadón, A. *et al*. Giant and Anisotropic Enhancement of Spin-Charge Conversion in Graphene-Based Quantum System. *Adv. Mater*. **37**, 2418541 (2025).

56. Tserkovnyak, Y., Brataas, A. & Bauer, G. E. W. Spin pumping and magnetization dynamics in metallic multilayers. *Phys. Rev. B* **66**, 224403 (2002).

57. D'Yakonov, M. I. & Perel, V. I. Spin Orientation of Electrons Associated with the Interband Absorption of Light in Semiconductors. *J. Exp. Theor. Phys*. **33**, 1954 (1971).

58. Boross, P., Dóra, B., Kiss, A. & Simon, F. A unified theory of spin-relaxation due to spin-orbit coupling in metals and semiconductors. *Sci. Rep*. **3**, 3233 (2013).

59. Safi, T. S. *et al*. Variable spin-charge conversion across metal-insulator transition. *Nat. Commun*. **11**, 476 (2020).

60. Malozemoff, A. P. Random-field model of exchange anisotropy at rough ferromagnetic-antiferromagnetic interfaces. *Phys. Rev. B* **35**, 3679–3682 (1987).

61. Nagaosa, N., Sinova, J., Onoda, S., MacDonald, A. H. & Ong, N. P. Anomalous Hall effect. *Rev. Mod. Phys*. **82**, 1539–1592 (2010).

62. Zhu, L., Zhu, L., Sui, M., Ralph, D. C. & Buhrman, R. A. Variation of the giant intrinsic spin Hall conductivity of Pt with carrier lifetime. *Sci. Adv.* **5**, eaav8025 (2019).

63. Wang, Y. *et al.* Enhancement of spintronic terahertz emission enabled by increasing Hall angle and interfacial skew scattering. *Commun. Phys.* **6**, 1–10 (2023).

64. Xu, X. *et al.* Giant Extrinsic Spin Hall Effect in Platinum-Titanium Oxide Nanocomposite Films. *Adv. Sci.* **9**, 2105726 (2022).

65. Kim, J. *et al.* Tuning Spin-Orbit Torques Across the Phase Transition in VO2/NiFe Heterostructure. *Adv. Funct. Mater.* **32**, 2111555 (2022).

66. Cheng, S. *et al.* Operando characterization of conductive filaments during resistive switching in Mott VO2. *Proc. Natl. Acad. Sci.* **118**, e2013676118 (2021).

67. Das, A. *et al.* Electric and magnetic tuning of Gilbert damping constant in LSMO/PMN-PT(011) heterostructure. *J. Phys. Condens. Matter* **35**, 285801 (2023).

68. Pati, S. P. & Taniyama, T. Voltage-driven strain-induced coexistence of both volatile and non-volatile interfacial magnetoelectric behaviors in LSMO/PMN-PT (0 0 1). *J. Phys. Appl. Phys.* **53**, 054003 (2019).

69. Sahoo, B. *et al.* High-Efficiency Continuous Spin-Conduction through NiO/Cu Bilayer Structure. *Nano Lett.* (2025) doi:10.1021/acs.nanolett.4c05923.

# Supplementary information for:

# Enhanced spin-to-charge conversion in epitaxial all-oxide La$_{0.67}$ Sr$_{0.33}$ MnO$_3$/NdNiO$_3$ bilayers at the nickelate metal-insulator phase transition.

Biswajit Sahoo*, Sarmistha Das*, Akilan K, Alexandre Pofelski, Sebastien Petit-Watelot, Juan-Carlos Rojas-Sánchez, Yimei Zhu, Alex Frano[m], and Eric E Fullerton[m].

## Contents

## Sample Fabrication:

12-nm-thick LSMO films were grown on NdGaO$_3$(*110*) (NGO) substrates by pulsed laser deposition. The substrate temperature during deposition was kept constant at 950$^0$C. The films were grown at an O$_2$ pressure of 200 mTorr. The films were then cooled down at the same pressure till 710$^0$C after which NNO was deposited at the same deposition pressure. The films were then annealed at the same temperature for 15 minutes at 7.6 Torr O$_2$ pressure and then cooled down to room temperature at the same pressure.

## X-Ray Diffraction:

The crystallinity of LSMO was determined via X-Ray Diffraction (XRD) using Rigaku Smartlab systems which employs Cu-Kα X-rays of wavelength 1.54 Å.  S1(a) shows the full XRD scan for LSMO(12) and LSMO(12)/NNO(16) samples. S1(b) shows the *220* peaks for LSMO(12)/NNO(4-16) samples showing an increase in strain with thinner NNO.

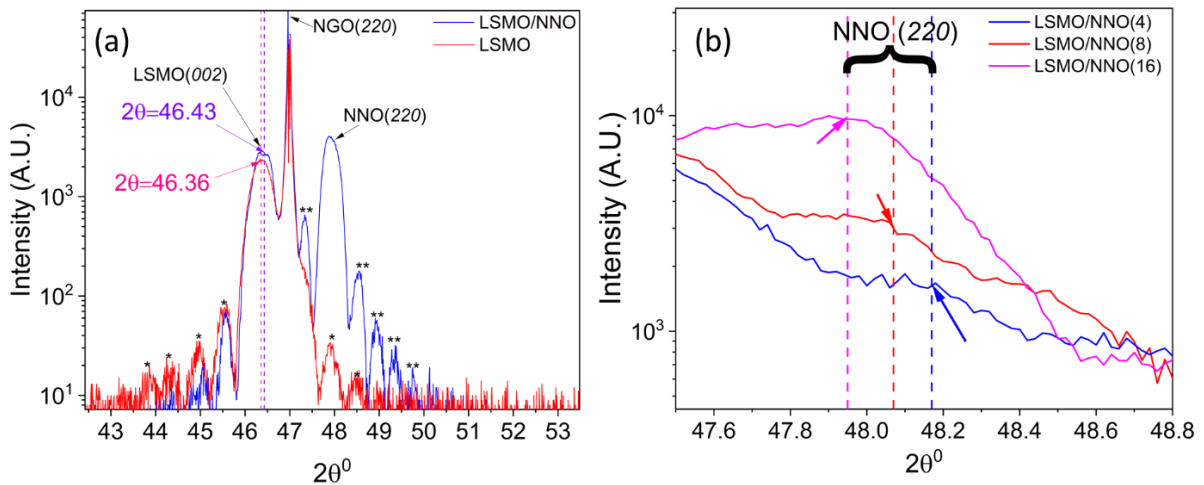

*Figure S4(a) XRD of LSMO(12) and LSMO(12)/NNO(16) bilayers showing the reduction of compressive strain of LSMO due to addition of NNO. (b) the NNO(220) peak for LSMO(12)/NNO(4-16) showing an increase in the tensile strain for thinner NNO.*

## STEM Characterization:

Samples were prepared using the in-situ lift-out method in the Helios G5 Dual Beam instrument operating at 30 keV. A final clean-up at 5 keV was performed to remove part of the amorphous material on the lamella. Samples were observed in a double aberration corrected JEOL ARM 200F cold FEG microscope operating at 200 keV. The STEM High Angle Annular Dark-Field (HAADF) imaging was done with a semi-convergence angle of 21 mrad and a current of 60 pA resulting in a probe size of approximately 80 pm.

The Electron Energy Loss Spectroscopy (EELS) analysis was performed using the GIF Continuum spectrometer and the K3 direct detection camera in Dual EELS counting mode using a dwell time of 50 ms (5 ms and 45 ms for the low loss and core loss part, respectively). The energy alignment was performed using the zero-loss peak. The acquired core loss EELS spectra were denoised using the Principal Component Analysis method (PCA), keeping the first 20 components. Figure S2 shows the EELS spectra of LSMO(20)/NNO(10) sample showing the chemically sharp LSMO/NNO interface.

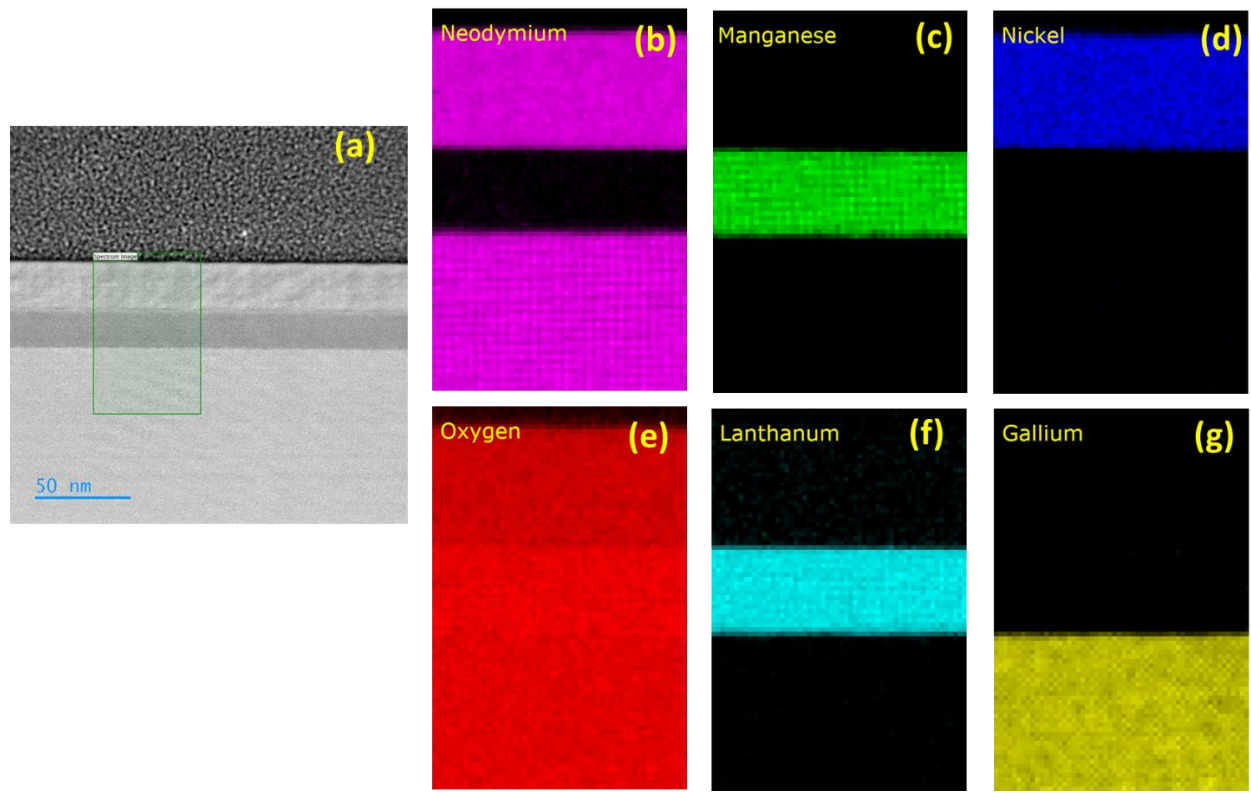

*Figure S5 TEM EELS spectra showing distribution of various elements in LSMO/NNO bilayer*

## Electrical Characterization:

The RT measurement of the samples were performed via the 4 probe Van der Pauw method from 300K to 50K at 2K intervals in the warming and cooling cycles to obtain the sheet resistance. The resistance was measured with 5 averages after each temperature point was stabilized.

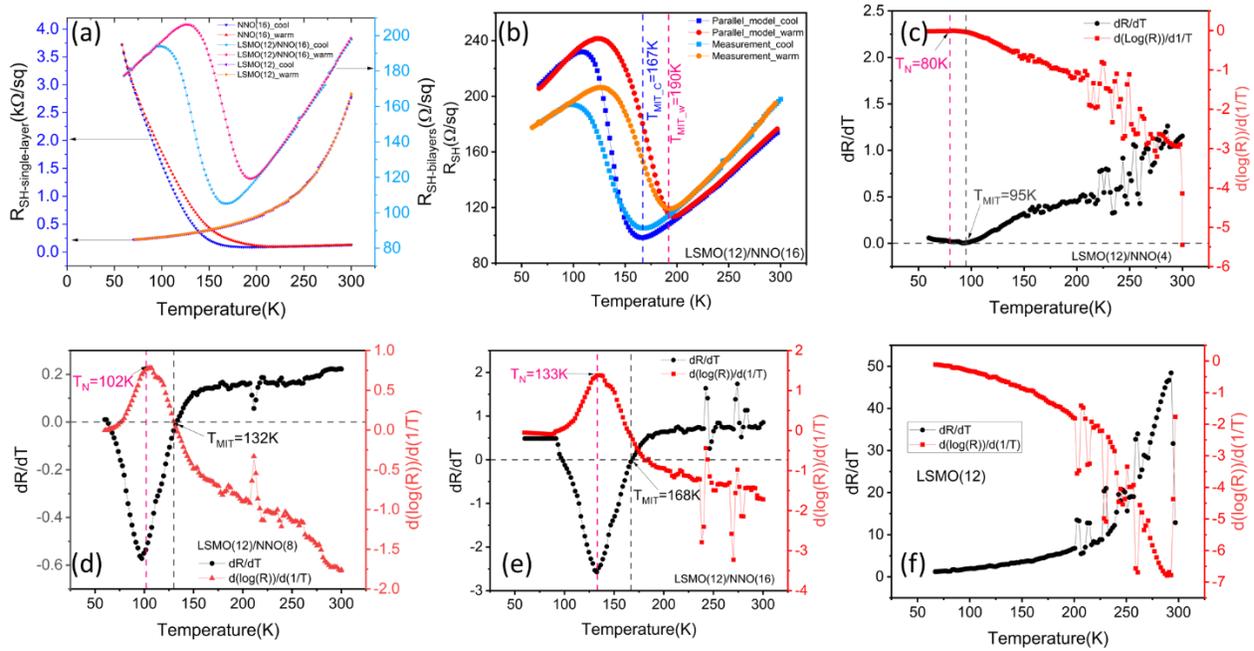

*Figure S6 (a) shows the R vs T for LSMO(12), NNO(16) and LSMO(12)/NNO(16). (b) shows the comparison between vs T of the measured and parallel resistor model ($1/R_{LSMO/NNO}=1/R_{LSMO}+1/R_{NNO}$) of LSMO(12)/NNO(16). (c)-(f) show the dR/dT and the d(log(R))/d(1/T) vs T showing the Neel and MIT temperature for LSMO(12)/NNO(4-16). LSMO(12) does not show any such transition.*

## Magnetometry measurements:

Temperature dependent magnetometry was performed using a vibrating sample magnetometer (VSM) by Quantum Design to confirm the ferromagnetic phase of the samples. FigureS4 shows some representative M vs H plots for NGO(*110*)/LSMO along the <001> and <1-10> orientations of the NGO substrate. Figure S5 shows the $M_s$ obtained from VSM for the LSMO(12) and LSMO(12)/NNO(8-16) samples showing an increase in the $T_C$ for the bilayer samples, possibly due to induction of strain on LSMO.

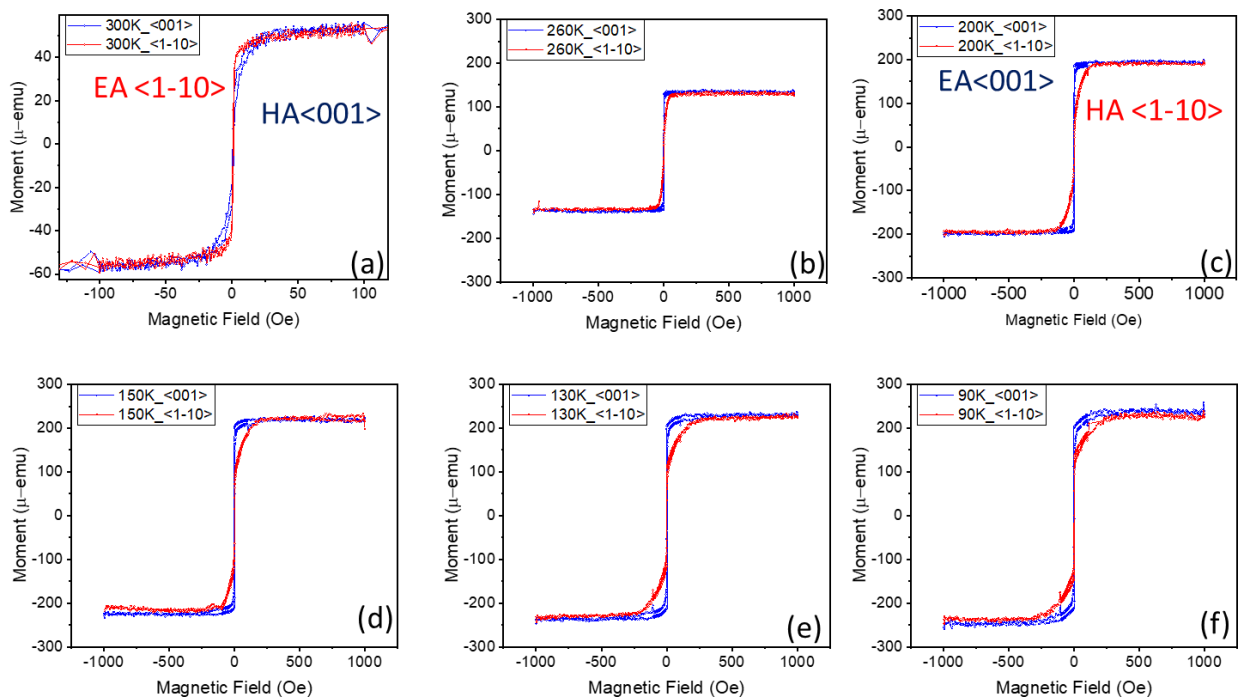

*Figure S7 MH loops for NGO(110)/LSMO(12) at different temperatures in the <001> and <1-10> direction of NGO substrate.*

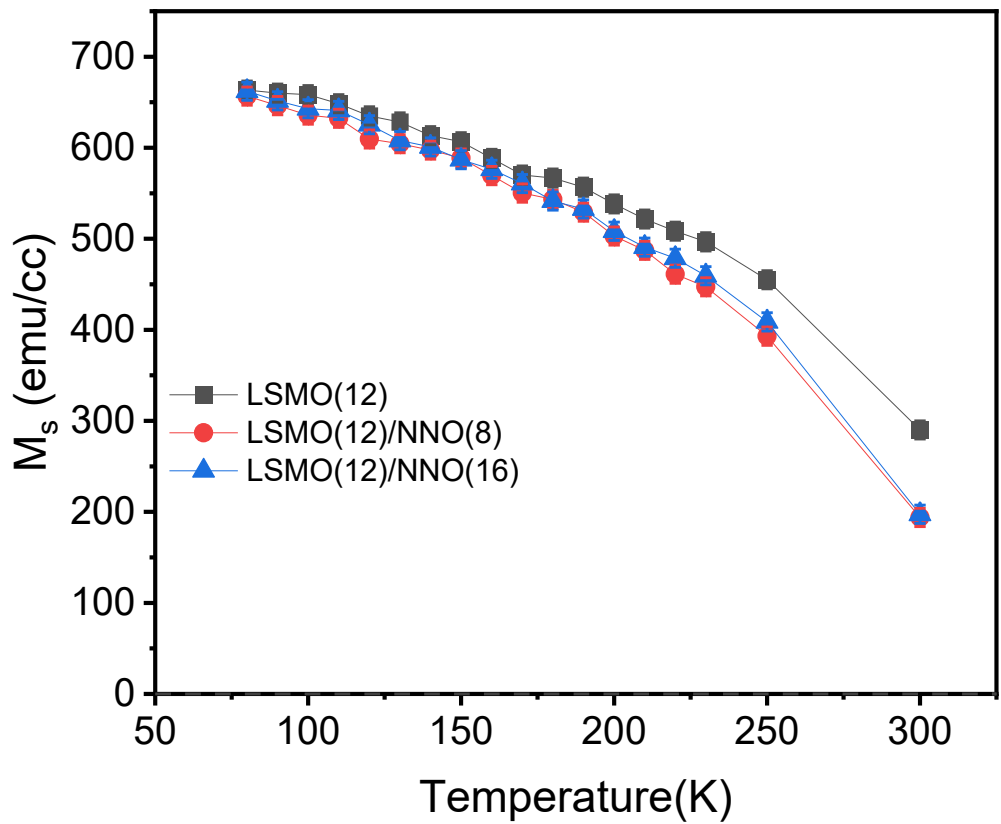

*Figure S8 Ms vs T for LSMO(12) and LSMO(12)/NNO(8-16)*

# Ferromagnetic Resonance:

We employ a field modulated flip chip coplanar waveguide FMR modified to adapt to a cryogenic chamber for temperature variation. We measure the spectra at different temperatures and subsequently evaluate the various magnetic parameters, more importantly, the Gilbert damping parameter as it provides direct insight into the interaction of magnetic moments of LSMO with NNO via dissipation of angular momentum from LSMO . The representative FMR spectra at 8GHz taken at various temperatures for LSMO(12) and LSMO(12)/NNO(16) are shown in Figure S6(a)-(b). By just qualitatively observing the lineshape of the spectra, we immediately see multiple modes at 300K, which is attributed to the lower exchange interaction strength of LSMO, as we are close to the paramagnetic phase transition temperature (≈360K)[1]. At lower temperatures, the signal strength increases, and the modes coalesce into a single stronger mode due to the stronger exchange coupling of the LSMO spins[1]. Similar observation has been seen in other systems involving LSMO ([2,3]). These spectra are fit to the following form containing a symmetric ($S_1$) and anti-symmetric ($A_1$) Lorentzian to extract the resonant field ($H_{res}$) and the linewidth ($\Delta H$):

$$\frac{dP}{dH} = off + mH + S_1 \frac{\Delta H^2 - 4(H - H_{res})^2}{(4(H - H_{res})^2 + \Delta H^2)^2} - 4A_1 \frac{\Delta H(H - H_{res})}{(4(H - H_{res})^2 + \Delta H^2)^2} \tag{1}$$

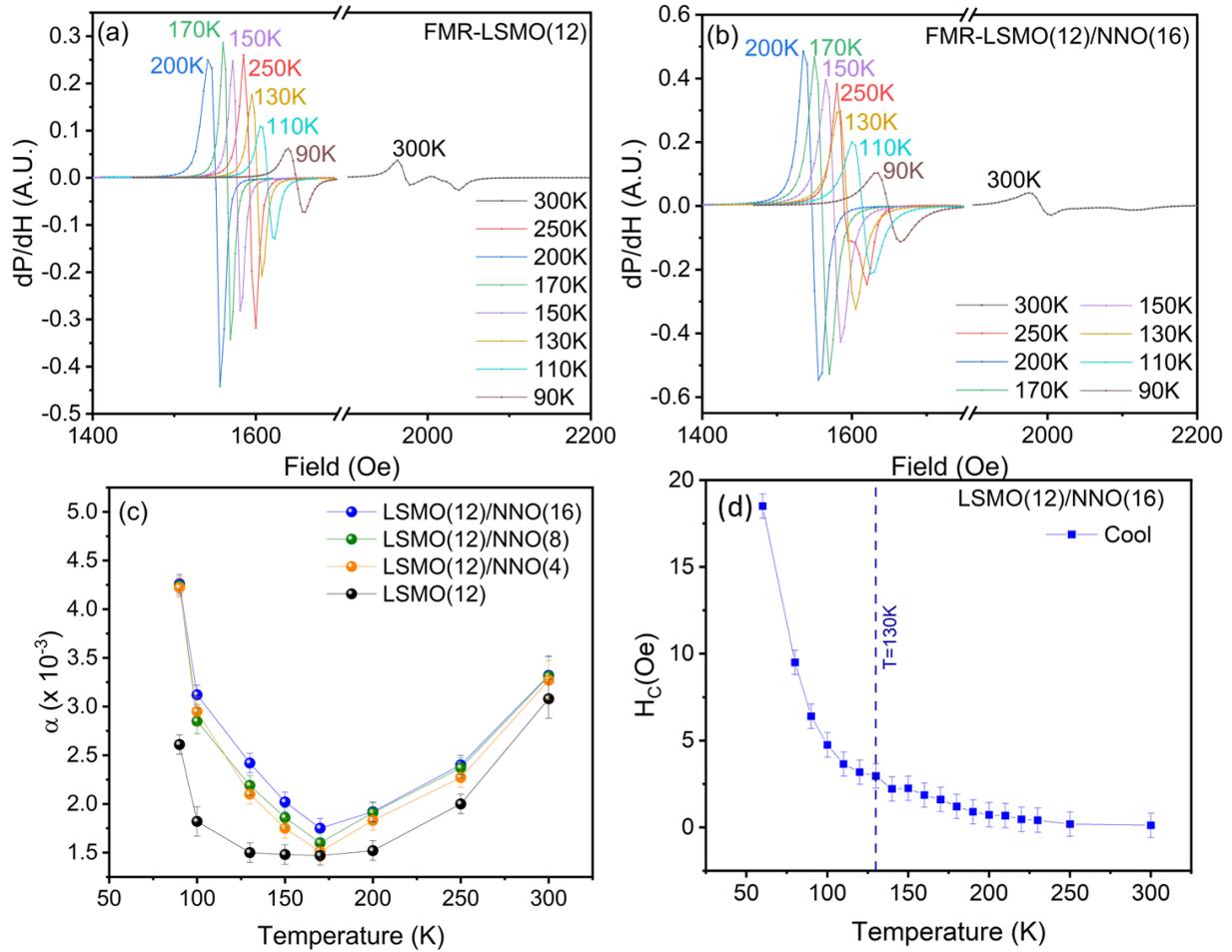

*Figure S9 (a) and (b) show the FMR spectra for LSMO(12) and LSMO(12)/NNO(16) respectively. The compilation of the Gilbert damping variation with temperature is shown in (c). (d) shows the variation of coercivity with temperature in LSMO(120/NNO(16) sample showing the onset of FM ordering.*

The variations of $H_{res}$ vs frequency ($f$) are fit to the Kittel equation[4]:

$$f = \frac{\gamma}{2\pi}\sqrt{(H_k + H_{res})(H_k + H_{res} + 4\pi M_{eff})} \qquad (2)$$

Here ϒ is the gyromagnetic ratio, $H_k$ is the in-plane uniaxial anisotropy field and $M_{eff}$ is the effective magnetization. Some representative fits for LSMO(12) in the <100> and <1-10> direction are shown in Fig S7(a)-(b). The $H_k$ values obtained by fits from Eq (2) reveal the increasingly negative values when measured along the <1-10> direction (Fig. S7(d)), indicating that the observed trend of $H_{Res}$ with temperature is driven by the anisotropy of LSMO (LSMO is slightly more strained in the <001> axis of NGO as opposed to <1-10>, subsequently inducing an in-plane anisotropy in LSMO creating a hard and an easy axis along the <001> and the <1-10> directions of NGO respectively, at room temperature [5]). To maintain consistency, measurements for all the samples have been performed along the <1-10> direction of NGO.

The variation of the linewidth (*ΔH*) with frequency provides us with the gilbert damping *α* of the system using the following equation:

$$\Delta H = \Delta H_0 + \frac{4\pi\alpha}{\gamma}f \qquad (3)$$

Here $\Delta H_0$ is the inhomogeneous linewidth broadening. The plots and fits to Eq 3 for LSMO(12)/NNO(4-16) are shown in Fig. S7. The fits follow a linear trend, indicating minimal contribution of effects such as two-magnon scattering and mosaicity broadening, which can introduce non-linearities in $\Delta H$ vs $f$ behavior. Lower value of α indicates lesser dissipation of the spin angular momentum (either into the lattice, or due to spin orbit effects), leading to better energy efficiency in spintronic devices.

At room temperature, the LSMO has a low damping of around 0.0030. The bilayer samples show a slightly increased damping values of around 0.0033 (Fig. 3c). The damping decreases, with a minima at 170K for all the samples, albeit with an increased damping value for the bilayer samples, which may be attributed to the additional spin dissipation in the NNO layer due to spin pumping. For T<170K the damping increases for all the samples, with a more prominent increase for the bilayers, as opposed to LSMO. Similar behavior has also been observed in NGO/LSMO/Pt where the Pt was deposited *ex-situ*[3], which was attributed to spin dissipation due to increased interaction between different Mn valence states in the bulk and at the top and bottom interface of NGO/LSMO/Pt. In the present case since the deposition is *in-situ*, we speculate that LSMO may have induced additional moment in the adjacent Ni ions in NNO due to magnetic proximity effects (MPE). The increased interaction between the Mn and Ni spins at lower temperatures may have caused the damping enhancement resulting in the similar trend of damping increase for T<170K in the bilayers. We attempted to quantify this induced moment, via $M_s$ vs T measurements for samples LSMO(12)/NNO(8-16). The bilayers exhibited a slightly lower Curie temperature ($T_C$) as opposed to LSMO (Fig S5), which may be attributed to the tensile strain [6] on LSMO caused by the overlying NNO layer, as exemplified by the LSMO(*002*) peaks of LSMO(12) and LSMO(12)/NNO(16) in Fig S1(a). However, we could not see any enhancement in $M_s$ due to MPE. This may indicate that the MPE is too weak to be observed via simple magnetometry, and thus we rely on literature[7,8] to explain this trend in damping due to MPE. In the present case, it is extremely difficult to disentangle the effect of MPE and spin pumping induced damping enhancement, and thus we assume that these two effects work in synergy. The AFM related damping enhancement contribution in the LSMO/NNO bilayers, however, is probably being smeared by the possible MPE + spin pumping at LSMO/NNO interface.

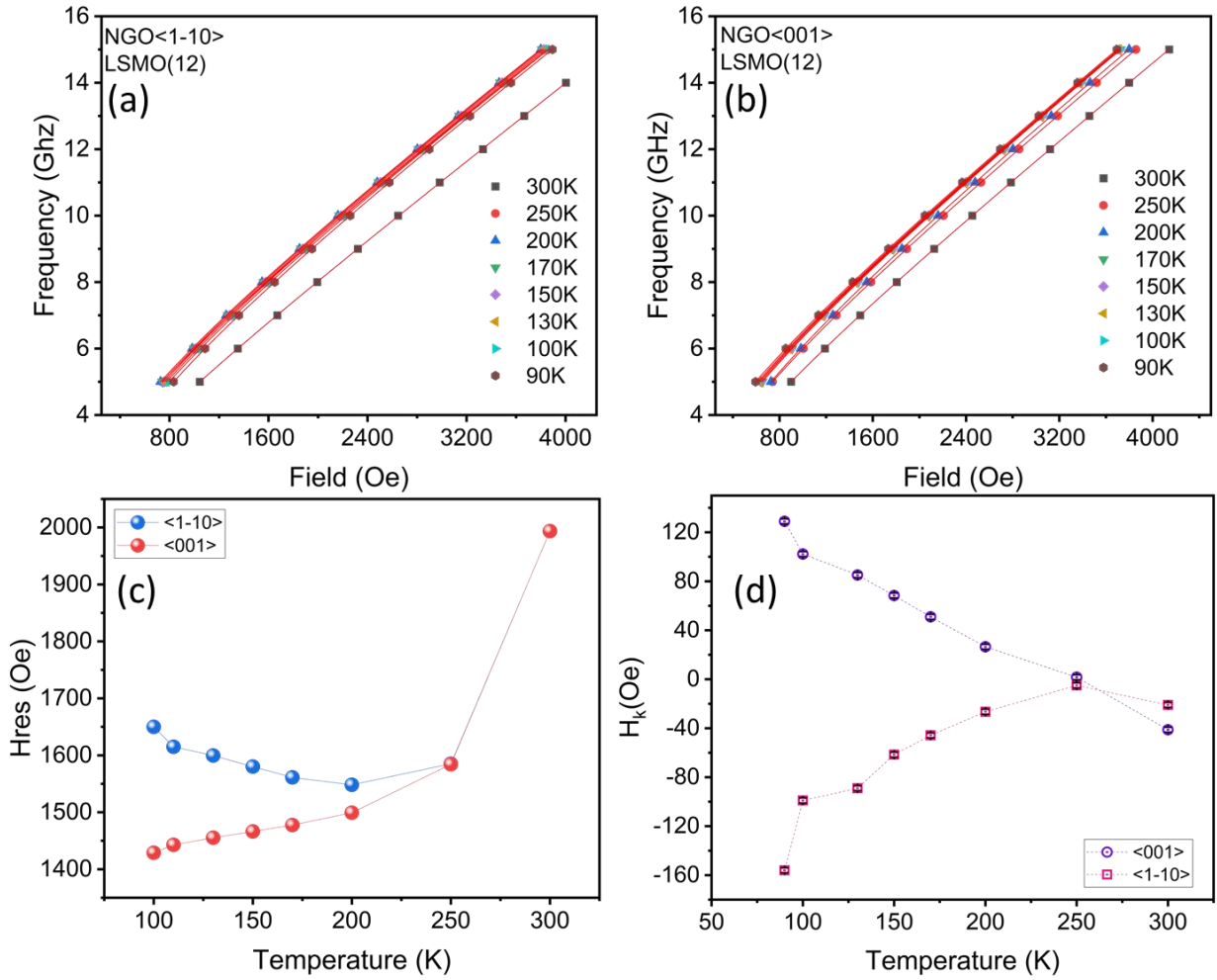

Figure S10 (a)-(b) Kittel fits for LSMO(12) in the <1-10> and <001> NGO directions.(c) shows the $H_{res}$ vs T at 8GHz. (d) shows the variation of $H_K$ obtained from Kittel fitting of (a) and (b).

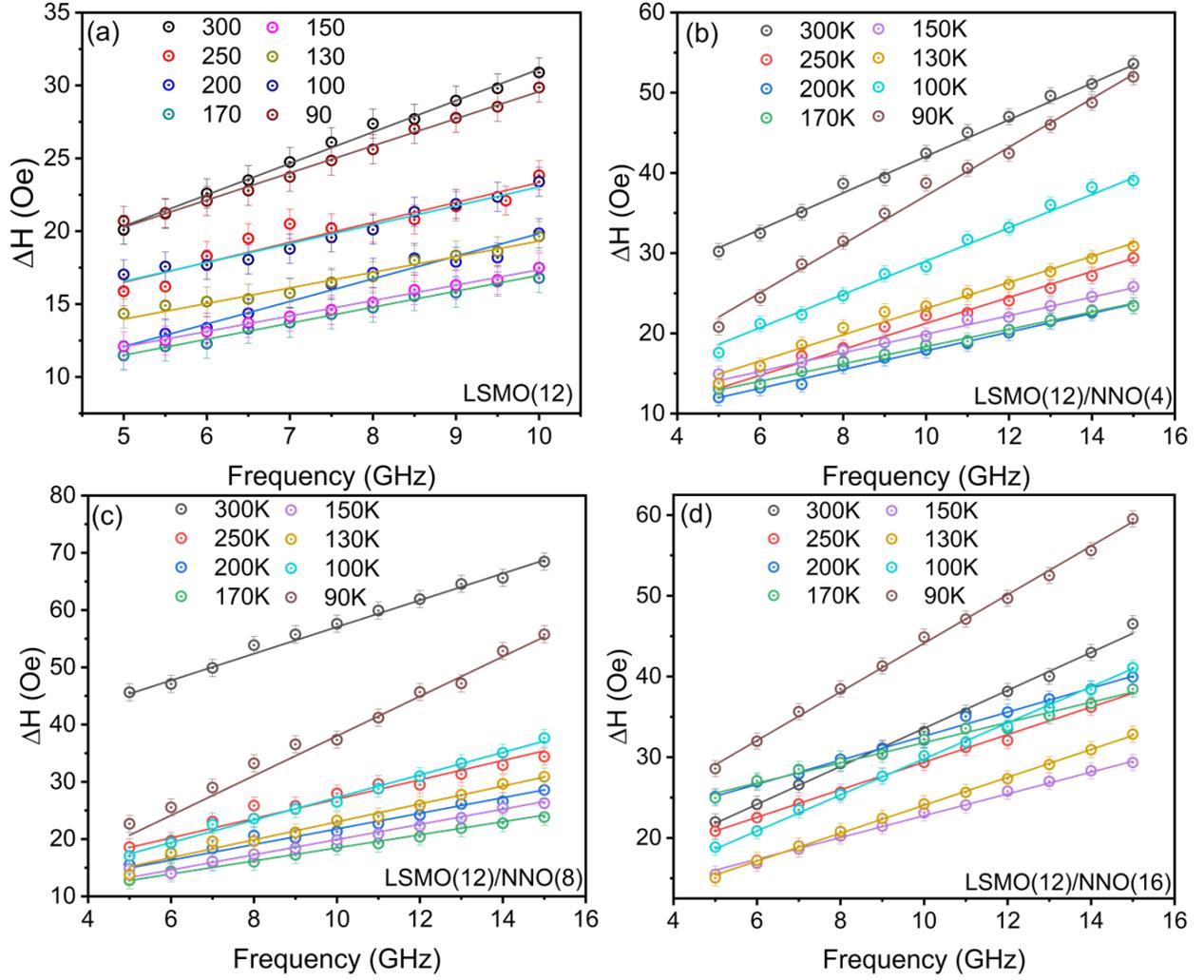

*Figure S11 Damping fits for LSMO(12)/NNO(0-16)*

To obtain the spin diffusion length ($\lambda_{SF}$), the variation in damping is fit to the following equation[9]:

$$\alpha = \alpha_{LSMO} + \frac{g\mu_B g_{Pd}^{\uparrow\downarrow}}{4\pi M_s t_{LSMO}}\left(1 - e^{-\frac{2t_{NNO}}{\lambda_{NNO}}}\right) \qquad (4)$$

Furthermore, to contrast with the spin diffusion length obtained by fitting of the $I_{C\text{-}SP}$ variation, the following function was used[10]:

$$I_{C-SP} = A * \lambda_{SF} \tanh\left(\frac{t_{NNO}}{2\lambda_{SF}}\right) \qquad (5)$$

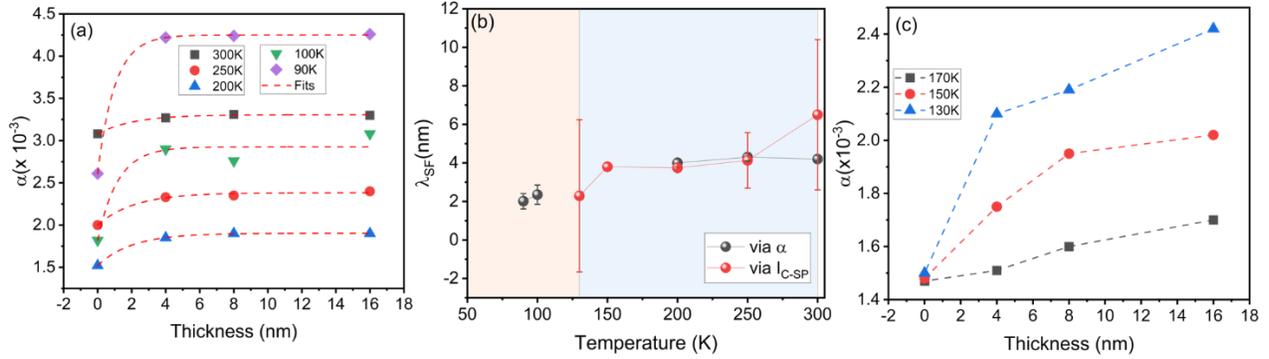

*Figure S12 Estimation of $\lambda_{SF}$ from damping variation (b) by the fits from Eq. 4 (a) (black spheres). Comparative values obtained by fitting the $I_{C-SP}$ to a tanh (Eq.5) function is shown in red spheres. Both the cases show good agreement in values wherever fitting was possible. (c) shows the variation of damping for LSMO(12)/NNO(0-16). from 170-150K for which the fitting did not work, possibly due to enhanced interaction of LSMO with the AFM domains of NNO in LSMO(12)/NNO(16).*

## Device fabrication:

For Fabricating SP-FMR devices, the film was patterned into devices of 500 x 10 µm wires via UV lithography and Ar ion etching. Next, 200 nm of $SiO_2$ was deposited on the entire substrate, leaving areas for gold contact deposition. Finally, 5-nm Ti and 150-nm Au layers were deposited by evaporation to fabricate ground-signal-ground waveguides and the contacts by lift-off. For measurements, the devices were placed in a cryo-chamber with 4 probes: two for sending RF currents and two for measuring the resulting voltage. An external field perpendicular to the RF current was swept, and the resulting voltage (due to spin rectification effects) was recorded by a nano-voltmeter.

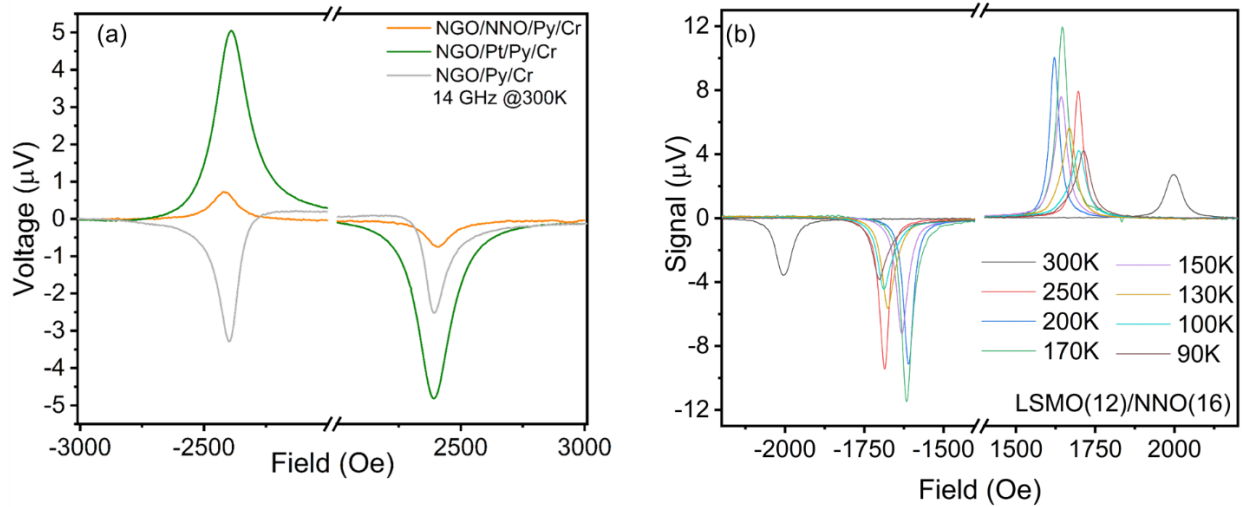

*Figure S13 SP-FMR spectra of (a) NNO(23)/Py(7)/Cr(3), Pt(5)/Py(7)/Cr(3) and (b) LSMO(12)/NNO(16) for all the measured temperatures.*

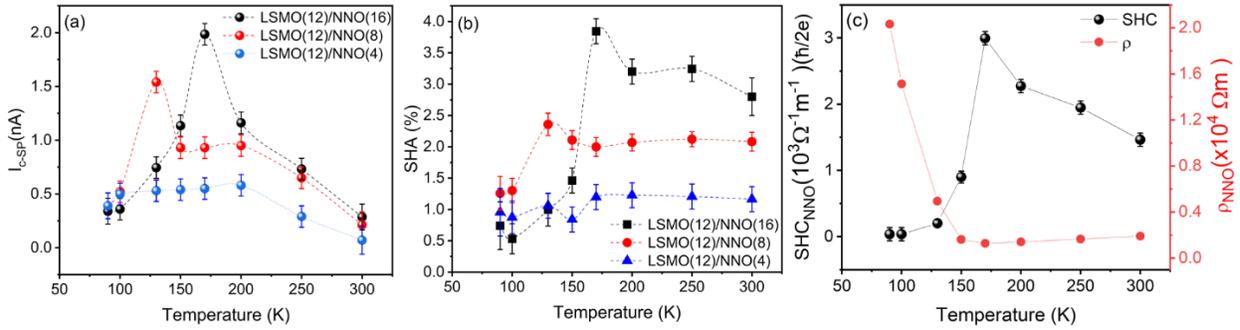

*Figure S14 Variation of SHA , $I_c$ and spin Hall conductivity with temperature.*

# References:

1.	Demidov, V. V. & Ovsyannikov, G. A. Temperature dependence of interlayer exchange interaction in $La_{0.7}Sr_{0.3}MnO_3$/$SrRuO_3$ heterostructure. *J. Appl. Phys.* **122**, 013902 (2017).

2.	Lee, H. K. *et al.* Magnetic anisotropy, damping, and interfacial spin transport in Pt/LSMO bilayers. *AIP Adv.* **6**, 055212 (2016).

3.	Sahoo, B. *et al.* Temperature Dependent Spin Dynamics in $La_{0.67}Sr_{0.33}MnO_3$/Pt Bilayers. *Adv. Mater. Interfaces* **n/a**, 2401038 (2025).

4.	Kittel, C. On the Theory of Ferromagnetic Resonance Absorption. *Phys. Rev.* **73**, 155–161 (1948).

5.	Boschker, H. *et al.* Strong uniaxial in-plane magnetic anisotropy of (001)- and (011)-oriented $La_{0.67}Sr_{0.33}MnO_3$ thin films on $NdGaO_3$ substrates. *Phys. Rev. B* **79**, 214425 (2009).

6.	Tsui, F., Smoak, M. C., Nath, T. K. & Eom, C. B. Strain-dependent magnetic phase diagram of epitaxial $La_{0.67}Sr_{0.33}MnO_3$ thin films. *Appl. Phys. Lett.* **76**, 2421–2423 (2000).

7.	Mattoni, G. *et al.* Striped nanoscale phase separation at the metal–insulator transition of heteroepitaxial nickelates. *Nat. Commun.* **7**, 13141 (2016).

8.	Preziosi, D. *et al.* Direct Mapping of Phase Separation across the Metal–Insulator Transition of $NdNiO_3$. *Nano Lett.* **18**, 2226–2232 (2018).

9.	Shaw, J. M., Nembach, H. T. & Silva, T. J. Determination of spin pumping as a source of linewidth in sputtered $Co_{90}Fe_{10}$/Pd multilayers by use of broadband ferromagnetic resonance spectroscopy. *Phys. Rev. B* **85**, 054412 (2012).

10.	Fache, T., Rojas-Sanchez, J. C., Badie, L., Mangin, S. & Petit-Watelot, S. Determination of spin Hall angle, spin mixing conductance, and spin diffusion length in CoFeB/Ir for spin-orbitronic devices. *Phys. Rev. B* **102**, 064425 (2020).